# A note on classification of proper teleparallel conformal symmetries of spherically symmetric static space-times in the teleparallel theory of gravitation using diagonal tetrads

Muhammad Amer Qureshi[a], Ghulam Shabbir[a,c] and K S Mahomed[b]

(a) Faculty of Engineering Sciences, GIK Institute of Engineering Sciences and Technology, Topi, Swabi, KPK, Pakistan. [c]Email: shabbir@giki.edu.pk

(b) School of Computer Science and Applied Mathematics, University of Witwatersrand, Johannesburg, Wits 2050, South Africa

**Abstract**

A study of proper teleparallel conformal vector field in spherically symmetric static space-times is given using the direct integration technique and diagonal tetrads. In this study we show that the above space-times do not admit proper teleparallel conformal vector fields.

## 1. Introduction

Gravitation is got behind by the curvature of space-times in the theory of general relativity. Particles are compelled by the curvature of space-times to follow the geodesics. Thus the theory of general relativity is actually geometrized by the gravitation. This theory might have some limitations at classical level. The predictions of general relativity are in accordance to the experimental data available so far. The unification of fundamental forces has remained a favorite topic in Physics. Due to uncertain behavior of general relativity at quantum level, the researchers needed the unified theory at this stage. In order to get a unified theory for general relativity and quantum theory, Einstein made an attempt in which he did not succeed. Here, we study his proposed theory, known as teleparallel theory of gravity as an alternate theory of gravitation without unifying it with quantum theory. Our work is confined for proper conformal symmetries of spherically symmetric static space-times in the teleparallel theory of gravitations. Teleparallel theory of gravity is based on Weitzenböck geometry [1]. In teleparallel theory torsion acts as a force on objects. Thus in teleparallel theory of gravitation there are no geodesics but there are force equations [2]. The general relativity is well described mathematically by Einstein's field equations, in which one side tells us the geometry of the space-time and the other side gives the physics of the space-time. Solutions of these highly nonlinear equations require certain symmetry restrictions on the space-time metric. The importance of the study of symmetries is quite clear in general relativity because the laws of conservation of the matter in the space-time can be studied and well understood with the help of these symmetry restrictions [3]. These symmetry restrictions not only give us the laws of conservation but also provide some geometrical features and physical information about the space-time. For example, in general relativity self-similarity solutions are extensively used for cosmological perturbations, star formation, gravitational collapse, primordial black holes, cosmological voids and cosmic censorship [4]. In [5] the authors introduced the teleparallel version of the Lie derivative for Killing vector fields and used those equations to find the teleparallel Killing vector fields in Einstein universe. In [6-13] the authors made further progress to classify different space-times according to their teleparallel Killing vector fields. From the above study they showed that some time teleparallel theory gives more conservation laws as compared with the theory of general relativity. In [14-17] the authors extended this work for proper teleparallel homothetic vector fields and they classified different space-times according to their proper teleparallel homothetic vector fields.

Keeping in view the importance of teleparallel theory the authors further extended this work to teleparallel conformal vector fields for different kind of space-times [18-22] and they found very special classes of the space-times admitted proper teleparallel conformal vector fields. In this paper we are interested to find the proper teleparallel conformal vector fields in spherically static space-times. The current study, in spherically symmetric space-times, will not only help to understand the geometrical and physical properties of the space-time but also to find the effect of torsion on the laws of gravitation. These results may give us interesting information about the compatibility of both the theories. The procedure for finding Weitzenböck connections and torsion components are given in detail in [17]. Teleparallel Killing equation is defined as [5]

$$\mathop{L}_{X}^{T} g_{\mu\nu} \equiv g_{\mu\nu,\rho}X^{\rho} + g_{\rho\nu}X^{\rho}{}_{,\mu} + g_{\mu\rho}X^{\rho}{}_{,\nu} + X^{\rho}(g_{\theta\nu}T^{\theta}{}_{\mu\rho} + g_{\mu\theta}T^{\theta}{}_{\nu\rho}) \text{ where } \mathop{L}_{X}^{T}$$

denotes the teleparallel Lie derivative with respect to the vector field $X$ and $T^{\theta}{}_{\nu\rho}$ denotes the torsion tensor which is anti-symmetric with respect to its lower two indices. Teleparallel conformal is defined as [18, 19]

$$\mathop{L}_{X}^{T} g_{\mu\nu} = 2\,\alpha\,g_{\mu\nu}, \tag{1}$$

where $\alpha$ is a function on $M$. If $\alpha$ becomes constant on $M$, then $X$ is called teleparallel homothetic (proper teleparallel homothetic when $\alpha \neq 0$) while if $\alpha = 0$ it becomes teleparallel Killing. Other wise $X$ is called proper teleparallel conformal vector field.

## 2. Main Results

Consider spherically symmetric static space-times in usual coordinate system $(t, r, \theta, \varphi)$ (labeled by $(x^0, x^1, x^2, x^3)$, respectively) with the line element [23]

$$ds^2 = -e^{A(r)}dt^2 + e^{B(r)}dr^2 + r^2(d\theta^2 + \sin^2\theta\,d\varphi^2), \tag{2}$$

where $A(r)$ and $B(r)$ are functions of $r$ only. For above mentioned space-time (13) we can write the non-trivial tetrad components $S_a^{\mu}$ and its inverse non trivial tetrad components $S_\mu^a$ [24]

$$S_a^{\mu} = \text{diag}(e^{A(r)}, e^{B(r)}, r^2, r^2\sin^2\theta),\ S_\mu^a = \text{diag}(e^{-A(r)}, e^{-B(r)}, r^{-2}, r^{-2}\sin^{-2}\theta). \tag{3}$$

The corresponding non-zero torsion components can be obtained as [17]

$$T^0{}_{10} = \frac{A'(r)}{2},\ T^2{}_{12} = \frac{1}{r},\ T^3{}_{13} = \frac{1}{r}. \tag{4}$$

A vector field $X$ is called teleparallel conformal vector field if it satisfied equation (1). Expanding equation (1) and using equations (2) and (4) one has

$$X^0{}_{,0} = X^2{}_{,2} = \alpha, \tag{5}$$

$$2X^1{}_{,1} + B'(r)X^1 = 2\alpha, \tag{6}$$

$$\sin^2\theta\,X^3{}_{,2} + X^2{}_{,3} = 0, \tag{7}$$

$$2e^{B(r)} X^1{}_{,0} - 2e^{A(r)} X^0{}_{,1} - A'(r)e^{A(r)} X^0 = 0, \tag{8}$$

$$e^{B(r)} X^1{}_{,2} + r^2 X^2{}_{,1} + r X^2 = 0, \tag{9}$$

$$e^{B(r)} X^1{}_{,3} + r^2 \sin^2\theta X^3{}_{,1} + r \sin^2\theta X^3 = 0, \tag{10}$$

$$r^2 X^2{}_{,0} - e^{A(r)} X^0{}_{,2} = 0, \tag{11}$$

$$r^2 \sin^2\theta X^3{}_{,0} - e^{A(r)} X^0{}_{,3} = 0, \tag{12}$$

$$\cot\theta X^2 + X^3{}_{,3} = \alpha, \tag{13}$$

where $\alpha = \alpha(r)$. Solving equations (5), (6) and (7) one gets

$$X^0 = \alpha t + P^1(r,\theta,\varphi), \quad X^1 = e^{-\frac{B(r)}{2}} \int e^{\frac{B(r)}{2}} \alpha\, dr + e^{-\frac{B(r)}{2}} P^2(t,\theta,\varphi), \tag{14}$$
$$X^2 = \alpha\theta + P^3(t,r,\varphi), \quad X^3 = \cot\theta\, P^3_\varphi(t,r,\varphi) + P^4(t,r,\varphi),$$

where $P^1(r,\theta,\varphi), P^2(t,\theta,\varphi), P^3(t,r,\varphi)$ and $P^4(t,r,\varphi)$ are functions of integration. Now we are interested to find the unknown values of $P^1(r,\theta,\varphi), P^2(t,\theta,\varphi), P^3(t,r,\varphi)$ and $P^4(t,r,\varphi)$ using remaining six equations. If one proceeds further after some tedious and lengthy calculations one finds that

$$X^0 = \alpha t + \theta K^2(r) + K^3(r), \quad X^1 = e^{-\frac{B(r)}{2}} \int e^{\frac{B(r)}{2}} \alpha\, dr + e^{-\frac{B(r)}{2}} P^2(t,\theta,\varphi), \tag{15}$$
$$X^2 = \alpha\theta + tK^1(r) + D^2(r,\varphi), \quad X^3 = \cot\theta\, D^2_\varphi(r,\varphi) + D^3(r,\varphi),$$

where $K^1(r)$, $K^2(r)$, $K^3(r)$, $D^2(r,\varphi)$ and $D^3(r,\varphi)$ are functions of integration. Differentiating equation (13) with respect to $t$, and using equation (15) we get $K^1(r) = 0$, plugging back in the same equation and differentiating with respect to $\theta$ twice, we obtain $\sin^2\theta\, \alpha = 0 \Rightarrow \alpha = 0$. Hence no proper teleparallel conformal vector fields exist. It also follows from [17] that the above space-times do not admit proper teleparallel homothetic vector fields. Here for the above space-times teleparallel conformal vector fields are teleparallel Killing vector fields.

**Summary**

In this paper we classify spherically symmetric static space-times according to their proper teleparallel conformal vector fields in teleparallel theory of gravitation using direct integration technique and diagonal tetrads. We have shown that spherically symmetric static space-times do not admit proper teleparallel conformal vector fields in teleparallel theory of gravitation. The teleparallel conformal vector fields are teleparallel Killing vector fields.